\documentclass[aps,pre,twocolumn,superscriptaddress,showkeys,showpacs]{revtex4}
\usepackage{graphicx, epsfig}
\usepackage{epstopdf}
\usepackage{amssymb, amsmath, amsfonts, multirow, rotate, color}

\DeclareGraphicsExtensions{.  jpg, .  pdf,  .  mps,  . png,  . eps,  .  ps,  .  EPS}
\usepackage{amsmath}
\usepackage{color}

\def\be{\begin{equation}}
\def\ee{\end{equation}}

\def\bc{\begin{center}}
\def\ec{\end{center}}
\def\bea{\begin{eqnarray}}
\def\eea{\end{eqnarray}}
\newcommand{\avg}[1]{\langle{#1}\rangle}
\newcommand{\Avg}[1]{\left\langle{#1}\right\rangle}
\begin{document}

\title{Dense Power-law Networks and Simplicial Complexes}

\author{Owen T. Courtney}
\affiliation{School of Mathematical Sciences, Queen Mary University of London, E1 4NS, London, UK}
\author{Ginestra Bianconi}
\affiliation{School of Mathematical Sciences, Queen Mary University of London, E1 4NS, London, UK}


\date{Received: date / Accepted: date}

\begin{abstract}
There is increasing evidence that dense networks occur in on-line social networks, recommendation networks and in the brain. In addition to being dense, these networks are often also scale-free, i.e. their degree distributions follow $P(k)\propto k^{-\gamma}$ with $\gamma\in(1,2]$. Models of growing networks have been successfully employed to produce scale-free networks using preferential attachment, however these models can only produce sparse networks as the numbers of links and nodes being added at each time-step is constant. Here we present a modelling framework which produces networks that are both dense and scale-free. The mechanism by which the networks grow in this model is based on the Pitman-Yor process.
Variations on the model are able to produce undirected scale-free networks with exponent $\gamma=2$ or directed networks with power-law out-degree distribution with tunable exponent $\gamma \in (1,2)$.
We also extend the model to that of directed  $2$-dimensional simplicial complexes. Simplicial complexes are   generalization of networks that can encode the many body interactions between the parts of a complex system and as such are becoming  increasingly popular to characterize different datasets ranging from social interacting systems to the brain.  Our model produces dense directed simplicial complexes with power-law distribution of the generalized out-degrees of the nodes.

\end{abstract}

\maketitle
\section{Introduction}

Networks \cite{NS,Newman_book,Doro_book} are a powerful framework to characterize the structure and the function of complex systems as diverse as  social networks, technological networks, molecular networks and the brain. 
Power-law degree distributions $P(k)=Ck^{-\gamma}$ have been identified as a universal property  of complex networks \cite{BA,NS} strongly affecting their dynamical behaviour \cite{Doro_book}.

The pivotal Barab\'asi and Albert model \cite{BA} has explained  the emergence of power-law distributions in complex networks by including two simple yet fundamental elements: growth and preferential attachment.  This pivotal work has triggered the formulation of several other models including the Bianconi-Barab\'asi model \cite{Fitness,BB}, the non-linear preferential attachment model \cite{Redner} and the model with initial attractiveness of the nodes \cite{Doro_exact_BA}. However, these models usually  assume that growth only occurs through the constant addition of nodes and links implying that these growing network models can only generate {\em sparse scale-free networks} with an average degree that does not depend on the network size. Therefore the emergent power-law network topologies  are characterized by  power-law exponents $\gamma\in (2,\infty)$. A model which uses preferential attachment to produce scale-free networks and which is also capable of producing dense networks is the duplication model \cite{Redner_dup1,Redner_dup2,Duplication1,Lambiotte}. In this model a new node is introduced at each time step and attaches to a randomly selected existing node of the network, as well as to each of the existing nodes' neighbors with a probability $p$. For $p<\frac{1}{2}$ it has been shown that the asymptotic growth in the number of links is linear and furthermore that the networks produced are sparse scale-free networks with $\gamma\in (2,\infty)$. For $p\geq \frac{1}{2}$ the networks are indeed dense, however the degree distribution in this regime is no longer scale-free \cite{Redner_dup1,Redner_dup2,Duplication1,Lambiotte}.

While the vast majority of complex networks are described by sparse networks, there is increasing evidence that networks with a diverging average degree, also called {\em dense networks} often occur in on-line social networks \cite{Zhang,Bianconi_dense1}, recommendation networks \cite{Zhang} and in the brain \cite{Bonifazi}.

In particular the vast majority of these networks are both dense and scale-free, i.e. they have a power-law degree distribution $P(k)\simeq k^{-\gamma}$ with power-law exponent $\gamma\in (1,2]$. Therefore it is rather relevant to develop new theoretical frameworks for modelling these networks. 

Dense network models are popular among statisticians and include  the graphon \cite{Diaconis,Chayes,Wolfe} in which the average degree increases linearly with the network size $N$, i.e. $\avg{k}=O(N)$ edge exchangeable models \cite{edge_model} and more diluted networks \cite{Caron_Fox}. However in the physics community dense networks are much less popular and have been treated only by a few authors.

The trouble for physicists with dense scale-free networks has been also enhanced after the publication of a work \cite{Gross} that shows that dense scale-free networks are not graphical, i.e. for the wide majority of degree distributions, if the cutoff is not chosen carefully it is not possible to construct a network without any  multiedge or any tadpole (a self-connected node, also known as a loop) that displays the desired degree distribution.

This result is certainly valid, and points to the fact that realizing dense power-law networks is more challenging than realizing sparse networks. However it would be misleading to claim from these results that dense power-law networks do not exist. In fact it is sufficient to allow for some moderate level of multiedges or soft constraints on the degree of the nodes or to impose a structural cutoff for generating dense power-law networks. 

Interestingly, the fact that dense scale-free networks do exist is demonstrated by the existence of a few modelling frameworks that extend the configuration model to dense scale-free networks \cite{Bianconi_dense1} by imposing a suitable structural cutoff, or that generate dense power-law networks with specific values of the power-law exponents (i.e. $\gamma=1$ \cite{Doro_gamma1} or $\gamma=1.5$ \cite{Bianconi_dense1}). 

Here we propose a theoretical framework that is designed to generate dense growing power-law networks with preferential attachment without imposing any ad hoc cutoff. We have based our approach on the pivotal Pitman-Yor process \cite{PitmanYor1,Gnedin,SMCRP} also known as the Chinese Restaurant Process. This process is originally defined for generating exchangeable partitions or, in more physical terms it is defined as a ball-in-the box process.
Our original aim is to generate growing dense power-law networks using a variation of the Pitman-Yor algorithm.
The first model that we have proposed is a dense undirected power-law network. In order to adapt the Pitman-Yor process to our purpose we have assumed that in the growing process multiedges are allowed giving rise to an undirected weighted network.

Interestingly we have found that this model is able to generate  dense power-law networks, however these networks are only marginally dense because they display a power-law exponent $\gamma=2$ which does not change by modifying the parameter of the model.
Therefore in this respect our work shows that realizing dense scale-free networks might be difficult, supporting the results found in Ref. \cite{Gross}.

We have therefore generalized the model by including a direction of the links showing that in this way it is possible to give rise to denser networks. However only the out-degree of these networks is scale-free while the in-degree is an homogeneous distribution.

Finally we have expanded this model to include dense simplicial complexes formed not only by nodes and links but also by triangles. Simplicial complexes \cite{Bassett,perspective,Emergent, Owen1,Krioukov,NGF,Hyperbolic,Owen2,Costa,Petri,BlueBrain}  are higher order networks that can be used for analysing collaboration networks, protein interaction networks or brain function.
From the network modelling perspective they provide a clear short-cut to generate networks with high clustering coefficients.

Our theoretical framework shows that dense simplicial complexes displaying dense power-law distribution of their generalized degree can be generated using a suitable modification of the Pitman-Yor process.
 
The paper is structured as follows: in Sec. \ref{sec:N&SC} we characterize the structure of networks in terms of the degrees and strengths of their nodes, and show how these concepts can be extended to simplicial complexes via the generalized degrees and generalized strengths. In Sec. \ref{sec:PY} we give an overview of the Yule-Simon and Pitman-Yor processes for generating power-law distributions with exponents $\gamma\in(2,\infty)$ and $\gamma\in (1,2]$ respectively. In Sec. \ref{sec:model} we present a modelling framework which exploits the Pitman-Yor process in order to produce dense scale-free networks and simplicial complexes. In Sec. \ref{sec:strengths} we derive mean-field expressions for the total number of nodes and their strengths. We use these results in Sec. \ref{sec:reinforce} to find the probability that a link or triangle is reinforced in any given time-step. In Sec. \ref{sec:degrees} we derive mean-field equations for the degrees, and show that the degree distributions are scale-free with dense exponent $\gamma\in (1,2]$. In Sec. \ref{sec:SvsK} we explore the relation between the strengths and degrees of the nodes. In Sec. \ref{sec:C&A} we examine the clustering and degree correlations produced by the model. Finally, in Sec. \ref{sec:conclusions} we give our conclusions.

\section{Networks and simplicial complexes}\label{sec:N&SC}

In this work we will model dense networks and simplicial complexes.

Networks describe the pairwise interactions between the elements of a complex system. This type of interaction can optionally be weighted and/or directed. 
An undirected network has a power-law degree distribution $P(k)$ if for large values of $k$,
\bea
P(k)\simeq Ck^{-\gamma}.
\eea 
Sparse networks are networks in which $\avg{k}$ does not diverge with the maximum degree $K$ of the network. Therefore sparse power-law networks have power-law exponent $\gamma\in(2,\infty)$. On the contrary dense networks are networks for which the average degree diverges with $K$. These networks have power-law exponents $\gamma\in (1,2]$.
For a weighted network it is also interesting to consider the strength distribution $P(s)$. The strength of a node is the sum of the weights of all of its links. Interestingly many networks such as airport networks and collaboration networks \cite{Barrat} are characterized by a scale-free strength distribution
\bea
P(s)\simeq \hat{C}s^{-\delta}
\eea 
with $\delta\in (1,\infty)$.

For directed networks it is possible to distinguish between the in-degree distribution and the out-degree distribution which can either be both power-law or one power-law and the other not.
Additionally we can distinguish also between the in-strength and the out-strength  indicating the sum of the weights of the in-coming and out-going links respectively. The in-strength and out-strength distributions characterize globally the properties of weighted directed networks. 

Simplicial complexes \cite{Owen1,Owen2,Emergent, NGF,Hyperbolic,Krioukov,Petri} are a generalization of networks that are able to encode the many-body interactions between more than two nodes. A $d$-dimensional simplex (also indicated as $d$-simplex) describes the interaction among a set of $d+1$ nodes.
For instance a $0$-simplex is a node, a $1$-simplex is a link, a $2$-simplex is a triangle and a $3$-simplex is a tetrahedron.
A $\delta$-face $\tilde{\alpha}$ of a $d$-simplex $\alpha$ is a $\delta$-simplex (with $\delta<d$) constructed from a proper subset of $\delta+1$ nodes of the simplex $\alpha$.
For instance the faces of a $2$-simplex are its $3$ links (3 $1$-simplices) and its $3$ nodes ($3$ $0$-simplices).

A  $d$-dimensional simplicial complex is  constructed  by gluing  simplices of  dimension smaller or equal to $d$ along their faces.
Additionally any simplicial complex satisfies the additional condition that if a  simplex belongs to the simplicial complex then all its faces must also belong to it.
We say a simplicial complex is `pure' if it is formed exclusively by $d$-dimensional simplices and their faces (i.e. the only simplices of dimension less than $d$ are those which are faces of a $d$-dimensional simplex). A network is a pure $1$-dimensional simplicial complex if it does not contain isolated nodes while a pure $2$-dimensional simplicial complex is constructed from triangles and the links and nodes belonging to the triangles.

We characterize the structure of simplicial complexes in terms of the generalized degrees and generalized strengths of their faces. The generalized degree $k_{d,\delta}(\alpha)$ of a $\delta$-dimensional simplex $\alpha$ is the number of simplices of dimension $d$ which $\alpha$ forms a subset of. If the simplices are weighted, then we define the generalized strength $s_{d,\delta}(\alpha)$ of a $\delta$-dimensional simplex $\alpha$ as the total weight of simplices of dimension $d$ which $\alpha$ forms a subset of.

\begin{figure} 
\includegraphics[width=0.5\columnwidth]{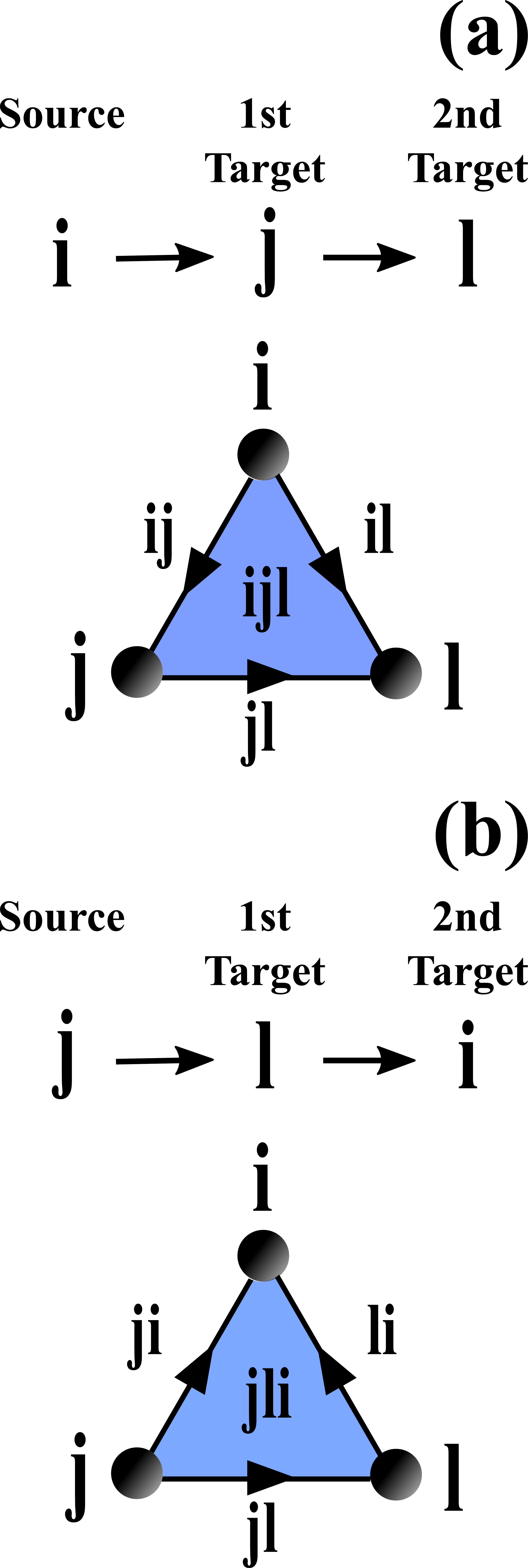}
\caption{(Color online). Diagram showing the relation between the direction of a triangle and the direction of its links. In panel (a) the triangle $ijl$ has node $i$ as its source node, node $j$ as its first target node and node $l$ as its second target node. There are three directed links present here: $(i,j)$, $(i,l)$ directed away from $i$ towards the two target nodes and $(j,l)$ directed from $j$ to $l$. The generalized out-strength and generalized out-degree are equal to $1$ for node $i$ and $0$ for nodes $j$ and $l$. In panel (b) the triangle $jli$ has node $j$ as its source node, node $l$ as its first target node and node $i$ as its second target node. The three directed links here are instead $(j,i)$, $(j,l)$ and $(l,i)$, while the generalized out-strength and generalized out-degree are now equal to $1$ for node $j$ and $0$ for nodes $i$ and $l$. }
\label{fig:triangles} 
\end{figure}

Here we will consider pure $2$-dimensional simplicial complexes which are both weighted and directional. Considering directed simplicial complexes is rather important for describing real world systems and has been proposed recently as a framework to capture the interplay between structure and dynamics of brain networks \cite{BlueBrain}. The triangles are directed in the sense that we map a differently `directed' or `oriented' triangle to each permutation of its three nodes, i.e. for three nodes labelled $i$, $j$ and $l$ we can create $6$ distinct directed triangles: $ijl$, $ilj$, $jil$, $jli$, $lij$ and $lji$. The first node in the triangle we call the `source' node, the second we call the `first target node' and the third we call the `second target node'. As with an undirected simplicial complex, the triangles contain their faces of dimension $1$ (links) and $0$ (nodes). In the simplicial complexes we present in this paper, these links are also directed, and their directions are determined by the direction of their parent triangle. The two links coming from the source node are directed outwards from the source node towards the two target nodes. The third link between the two target nodes is directed from the first target node towards the second target node. Figure \ref{fig:triangles} is a diagram showing the relation between the direction of a triangle and the direction of its links. 

Directed simplicial complexes allow us to distinguish between the simplices a node has `gained' from distinct attachment mechanisms through its 'generalized out-degree' and `generalized in-degree' (see Sec. IV.C). Of course the specific relation between the direction of a triangle and the directions of its links that we use in this paper is just a convention that we have chosen, and there are indeed other conventions that could be chosen instead. We have chosen ours as it produces simplices that have what could be called a `temporal direction', where the links of the simplices produced are acyclic. Interestingly these directed simplices have been recently used   \cite{BlueBrain} to analyse brain networks and coupling for topological information with the neuronal network dynamics.

\section{Dense scale-free distributions and the Pitman-Yor Process}\label{sec:PY}

The  Barab\'asi-Albert model \cite{BA} provides a fundamental mechanism for the emergence of  scale-free networks  with degree distribution $P(k)\sim k^{-\gamma}$ and diverging second moment $\avg{k^2}$. In the Barab\'asi-Albert model we start at time $t=1$ from a finite network and at each time $t>1$  we add one new node and $m$ links connected to the new node and to a node $i$ with degree $k_i$ chosen with probability 
\bea
\Pi_i^{BA}=\frac{k_i}{\sum_j k_j}.
\eea 
This probability implements the preferential attachment mechanism according to which  nodes which already have many links (i.e. have high degree $k_i$) are more likely to acquire new links.
The number of nodes $N(t)$ and the number of links $L(t)$ at time $t$ are deterministic variables growing linearly with time as we have $N(t)=t$ and $L(t)=mt$. Therefore the average degree of this network is independent on the network size and given by 
\bea
\Avg{k}=\frac{2L(t)}{N(t)}=2m.
\eea
The degree distribution $P(k)$ of the Barab\'asi-Albert model can be evaluated exactly in the large network limit $N(t)\gg1$ and is given by \cite{Redner,Doro_exact_BA}
\bea
P(k)=\frac{2\Gamma(m+2)}{\Gamma(m)}\frac{\Gamma(k)}{\Gamma(k+3)}\simeq 2m(m+1)k^{-3},
\eea 
where the latter approximated expression describes the tail of the distribution where  $k\gg1$.

Other variations of the Barab\'asi-Albert model have been shown to yield scale-free networks with tunable exponent $\gamma$.
However, considering network models in which the number of nodes and the number of links both increases linearly with time can only produce networks with power-law exponent $\gamma\in (2,\infty)$.
In fact power-law degree distributions $P(k)=Ck^{-\gamma}$ yield sparse network models (i.e. models in which the average degree is constant with the network size) only for $\gamma\in (2,\infty)$.

When we face the challenge of modelling dense power-law networks the question is whether it is possible to formulate alternative power-law network models in which the density of links increases in time resulting in a power-law degree distribution $P(k)\simeq Ck^{-\gamma}$ with power-law exponents $\gamma\in (1,2]$.

In order to formulate alternative models which include growth and preferential attachment but that generate dense power-law networks we have looked closely at the mathematical literature regarding stylized ball-in-the box models.

The mathematical origin of the power-law in the Barab\'asi-Albert model can be  rooted back to  a ball-in-the-box model called the Yule-Simon model \cite{Simon}. This is a discrete time stochastic process analogous to placing balls in a growing number of boxes with probabilities dependent on the number of balls already in the boxes. \\
The process starts at time $t=1$ with a single box with one ball in it. At each subsequent time step $t>1$ a new ball is introduced and is either placed in an existing box (with probability $\epsilon\in (0,1)$) or a new box is created (with probability $1-\epsilon$) and the ball is placed in it. The process may be thought of as producing a random partition of the set of balls introduced up to time $t$. For any given time $t$ we  indicate  the total number of boxes by $N$, and the total number of balls by $M$. Additionally  we indicate by $s_i$  the number of balls in the $i$th box. 
The reinforcement dynamics called preferential attachment in network models is implemented by assuming that the probability to place a new ball in the box $i$ grows linearly with the number of ball $s_i$ already in the box $i$.
Therefore in the Yule-Simon model the probability that the new ball is placed in box $i$ is
\bea
\Pi_i^{YS} = \left\{ \begin{array}{lll} \epsilon s_i/{t} & \mbox{for} &1< i\leq N, \\
(1-\epsilon) & \mbox{for} & i=N+1.
\end{array}\right.
\label{PiSimon}
\eea
Clearly in this model the average number of boxes $\Avg{N(t)}$ increases linearly with time, i.e. $\Avg{N(t)}=(1-\epsilon)t$. Moreover,  since at each time we add a new ball the number of balls at time $t$ is a deterministic variable given by $M(t)=t$. It follows that the average number of balls per box is constant in time, i.e.
\bea
\Avg{\frac{M}{N}}=\frac{1}{1-\epsilon}(1-\epsilon^t)\simeq\frac{1}{1-\epsilon}=O(1).
\eea
This implies that if the distribution $P(s)$ of balls in the boxes decays as a power-law $P(s)\simeq s^{-\gamma}$ it must necessarily have the power-law exponent $\gamma$ in the range $\gamma \in(2,\infty)$. In fact for  $\gamma\in (1,2]$ power-law distributions have a diverging average value.
The exact expression of the degree distribution for the Yule-Simon process can be calculated exactly in the limit $t\to \infty$ finding 
\bea
P(s)&=&\frac{1}{\epsilon}\Gamma\left(1+\frac{1}{\epsilon}\right)\frac{\Gamma(s)}{\Gamma(s + 1+1/\epsilon)}\nonumber \\
&\simeq& \frac{1}{\epsilon}\Gamma\left(1+\frac{1}{\epsilon}\right) s^{-\gamma}
\eea 
where  the last expression is derived in the limit $s\gg1$ and where the power-law exponent $\gamma$ is given by 
\bea
\gamma=1+\frac{1}{\epsilon}\in (2,\infty).
\eea
Therefore the Yule-Simon model using growth and preferential attachment can generate power-law distributions $P(s)\simeq Cs^{-\gamma}$ with $\gamma>2$ and having a finite average number of balls in the boxes.
The Barab\'asi-Albert model can be mapped to a balls-in-the-box model by assuming that each node corresponds to a box and each half-edge attached to a given node corresponds to a ball in the box.
Note that for the Barab\'asi-Albert model the number of nodes is a deterministic variable $N(t)=t$ as is the number of half edges, which is given by twice the number of links  $M(t)=2L(t)\simeq 2mt$. However the Barab\'asi-Albert model can be considered as being in the same {\em universality class} as the Yule-Simon process with $\epsilon=1/2$.

Interestingly, a different ball-in-the box model called the Pitman-Yor process \cite{PitmanYor1,Gnedin,SMCRP} is known to yield dense power-law distributions.
This model includes growth of the number of boxes, and reinforcement dynamics (preferential attachment) but enforces that the scaling of the number of balls with the number of boxes is superlinear.
In this way this model elegantly generates dense power-law distributions with exponent $\gamma\in (1,2]$. 
Starting at time $t=1$ with one ball in a single box at each time $t>1$ a new ball is added and either placed in an existing box $i$ or placed in a new box $i=N(t)+1$. Specifically the probability $\Pi_i^{PY}$ that the new ball goes in the box $i$ is parametrized by the parameter $\alpha\in (0,1)$ and given by 
\bea
\Pi_i^{PY} = \left\{ \begin{array}{lll} \frac{s_i (t) - \alpha}{t } & \mbox{for} &1< i\leq N, \\
 \frac{\alpha N }{t } & \mbox{for} & i=N+1.
\end{array}\right.
\label{PiPY}
\eea
As in the Yule-Simon process there is a growth in both the number of balls, and the number of boxes, with a preferential attachment mechanism for placing the balls. Unlike with Yule-Simon however, the probability of adding a new box is not constant, but depends on the number of boxes already added while decaying with the total number of balls already added \cite{PitmanYor1,SMCRP}. 
The marginal distribution of a single box in the large $t$ limit is \cite{PitmanYor1}
\bea
P(s)=\frac{\alpha}{\Gamma(1 - \alpha)} \frac{\Gamma(s - \alpha)}{\Gamma(s + 1)}\simeq \frac{\alpha}{\Gamma(1 - \alpha)} s^{-\gamma},
\eea 
where the latter expression is an approximation for the tail of the distribution (i.e. for $s\gg1$).
Here the power-law exponent $\gamma$ is given by 
\bea
\gamma=1+\alpha\in (1,2],
\eea
which implies that the distribution has a diverging first moment.
This is consistent with the fact that the number of balls increases superlinearly with the  expected number of boxes $\langle N(t)\rangle$ \cite{PitmanYor1} as we have 
\bea
\langle N(t)\rangle \simeq t^\alpha \nonumber \\
M(t)=t,
\eea
and therefore 
\bea
\Avg{\frac{M}{N}}={O}(t^{1-\alpha}).
\eea

In this paper we explore whether the Pitman-Yor process can be exploited in order to formulate dense power-law network models,  and we emphasize   the challenges posed by the density of the resulting networks.
In this endeavor our objective is to construct not only dense power-law networks formed by pairwise interactions but also dense simplicial complexes which allow one to go beyond the framework of pairwise interactions.

\section{Evolution of dense scale-free networks and simplicial complexes}\label{sec:model}

Here we introduce a modelling framework that exploits the Pitman-Yor process in order to generate dense weighted power-law networks and simplicial complexes.

Weighted dense power-law networks evolve by the subsequent addition of  nodes and the establishment of new links or reinforcement of already existing links. We consider both a version of the model where the links are directed and a version with undirected links. 

This modelling framework is then extended to  $2$-dimensional simplicial complexes, where the growth comes from the addition of nodes, links and triangles. These simplicial complexes have densely connected skeletons and power-law distributions of the generalized strengths and generalized degrees.

Unlike in many other models of growing networks, the number of nodes in the network at a given time is not a deterministic function of time but instead depends on the stochastic growth dynamics of the network. The relative probabilities of nodes being created or selected for reinforcement are analogous to a Pitman-Yor process, with an equivalent parameter $\alpha\in(0,1)$. Below we give the dynamics for each of the three versions of the model.

\subsection{Undirected network growth dynamics}
In the undirected network version of the model, we write the total number of nodes in the network at time $t$ as $N(t)$. Every pair of nodes $i,j\in \{1 , 2 , ... , N(t)\} $ has an associated weight $w_{ij}(t)$ taking non-negative integer values.\\
We start at $t=1$ with an undirected link between node $1$ and node $2$. 
At each time step $t\geq 1$  we select a node $i$ with probability  
\bea
\Pi_i^{U} = \left\{ \begin{array}{lll} \frac{s_i(t)-\alpha}{2t} & \mbox{for} &1\leq i\leq N \\
 \frac{\alpha N }{2t} & \mbox{for} & i=N+1.
\end{array}\right.
\label{PiU}
\eea 
We then update the value of $N$ and select a second node $j$ using the same algorithm. 
If the two selected nodes are not already linked we add a link between them, if they are already linked we reinforce the weight of the links. In other words the adjacency matrix element $a_{ij}$ and the weight $w_{ij}$ of the link $(i,j)$ are updated according to  
\bea
a_{ij}(t+1)&=&1, \nonumber \\
w_{ij}(t+1)&=&w_{ij}(t)+1.
\eea
Moreover, since the network is undirected we have 
\bea
a_{ji}(t)&=&a_{ij}(t),\nonumber \\
w_{ji}(t)&=&w_{ij}(t).
\eea
Therefore in this model we treat half-edges as the balls of the Pitman-Yor process and we treat the  nodes as the boxes of the Pitman-Yor process.
Therefore it is to be expected that the strength distribution will follow a power-law exponent with exponent $\gamma=1+\alpha$. However, given that the network is weighted, the degree distribution could potentially be significantly different because if a new link is placed between nodes that are already connected the strengths of these nodes will increase but their degrees will remain unchanged.

\subsection{Directed network growth dynamics}

The directed network growth dynamics assumes that links are directed and that only the source node of the links is chosen according to the Pitman-Yor reinforcement dynamics, while the target node is chosen uniformly at random among all the existing nodes of the network.
In this way we expect  that the density of links in the network will grow more rapidly than in the undirected case. In fact by choosing  the target node with uniform probability we are more likely to add new links because we are not biasing the target node to be a node of high degree.

In the directed version, we start at $t=1$ with a directed link from node $2$ to node $1$. \\
At each time step $t\geq 1$ a pair of nodes is selected and its weight is incremented by one. The source node is either an existing node or a new node, while the target node is chosen uniformly at random from the remaining existing nodes. The probability at time $t$ of selecting node $i$ as the source node is given by:
\bea
\Pi_i^{D}= \left\{ \begin{array}{lll} \frac{s_i^{out} (t) - \alpha}{t} & \mbox{for} &1< i\leq N \\
 \frac{\alpha (N-1) }{t} & \mbox{for} & i=N+1.
\end{array}\right.
\label{PiD}
\eea
The probability at time $t$ of selecting node $j$ as the target node is given by:
\bea
\hat{\Pi}_j^{N}=\frac{1}{N(t)}.
\eea
When both source node $i$ and target node $j$ have been selected we update the adjacency matrix element and weight of the link, i.e.
\bea
a_{ij}(t+1)&=&1, \nonumber \\
w_{ij}(t+1)&=&w_{ij}(t)+1.
\eea
Note that here we have chosen to select the source node of the link according to the Pitman-Yor dynamics while the target node is chosen with uniform probability. However it is also possible to consider a directed network model in which the target node is chosen according to the Pitman-Yor dynamics  and the source node is chosen uniformly at random. Since the two versions of the model are simply related by the inversion of the direction of the links here we omit the explicit treatment of the latter possible definition.

\subsection{Directed simplicial complex growth dynamics}

Here we consider a directed $2$-dimensional simplicial complex formed only by ``directed triangles".
The triangles are directed in the sense that each permutation of three nodes is associated with a different triangle. We say that the first node in the triangle is the ``source node", the second node is the ``first target node" and the third node is the ``second target node". For the triangles in this version of the model the links are also directed, and we have chosen the convention that the two links coming from the source node are directed away from the source node towards the target nodes and the third link is directed from the first target node to the second.
In this version of the model the triangles $ijl$ also have an associated weight $w_{ijl}(t)$ taking non-negative integer values. These weights are associated specifically to the directed triangles and thus are also directed in the same sense. We define the {\it generalized out-strength} $\tilde{s}^{out}_i$ of a node $i$ to be the total weight of triangles for which $i$ is the source node, i.e. $\tilde{s}^{out}_i = \sum_{j,l=1}^N w_{ijl}  $.

\begin{figure} 
\includegraphics[width=0.5\columnwidth]{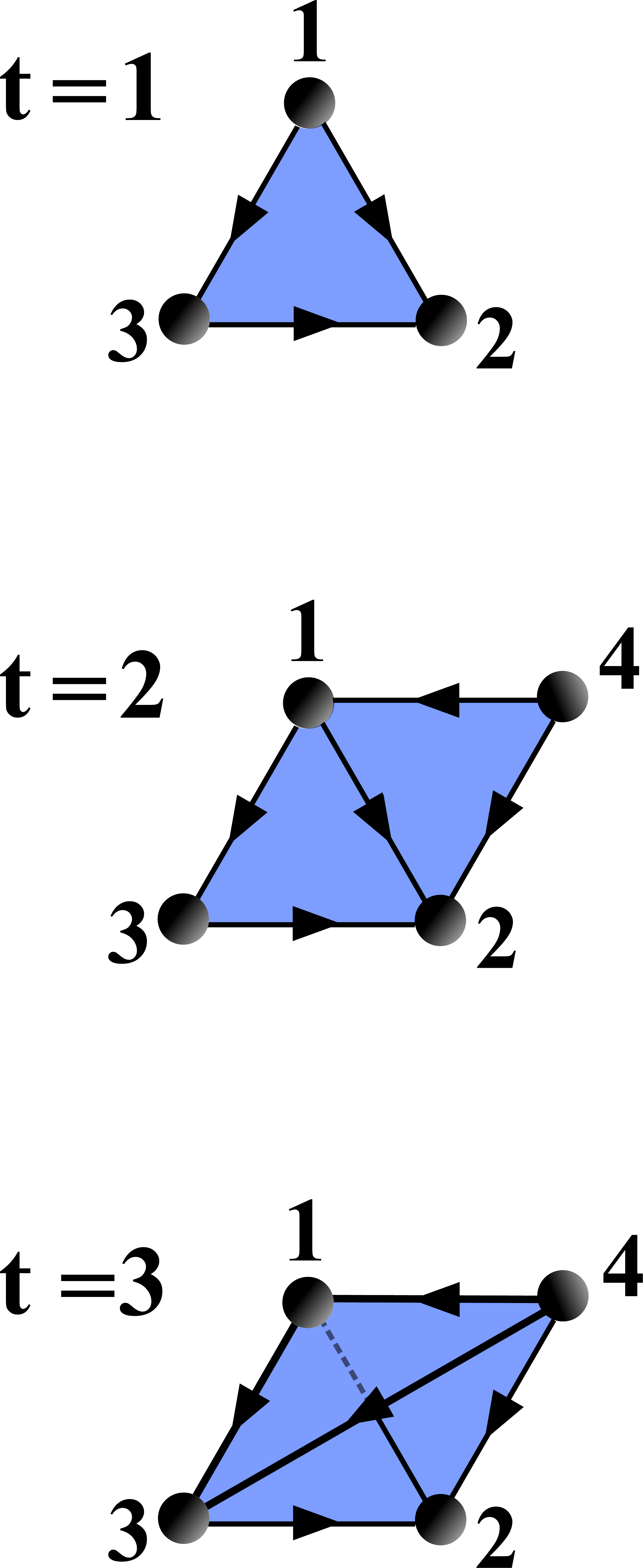}
\caption{(Color online). Diagram showing one possible way the simplicial complex could grow in the first three time-steps. At $t=1$ all simplicial complexes in the model start as the single triangle $123$. In this example, at $t=2$ a new node ($4$) is created and randomly selects the link $(1,2)$ to form the triangle $412$. At $t=3$ no new node is created. Instead, node $4$ is selected for reinforcement and randomly selects the link $(1,3)$ to form the triangle $413$. }
\label{fig:SCgrowth} 
\end{figure}

In this version of the model we start at $t=1$ with three nodes labelled $1,2,3$ and the single directed triangle $123$.
At each time step $t\geq 1$ we select a triangle to be created or reinforced. The source node $i$ of this triangle is selected with probability
\bea
\tilde{\Pi}_i^{SC} = \left\{ \begin{array}{lll} \frac{\tilde{s}^{out}_i(t)-\alpha}{t} & \mbox{for} &1\leq i\leq N, \\
 \frac{\alpha (N-2) }{t} & \mbox{for} & i=N+1.
\end{array}\right.
\label{PiSC}
\eea
Once this source node $i$ has been selected, a link $(j,l)$ is selected uniformly at random from the set of existing links with probability 
\bea
\hat{\Pi}_{jl}^{L}=\frac{1}{L(t)},
\eea
where $L(t)$ is the number of links in the simplicial complex at time $t$. If this triangle already exists then its weight is reinforced according to
\bea
w_{ijl}(t+1)=w_{ijl}(t)+1,
\eea 
while if it doesn't exist yet then it is created with initial weight one:
\bea
a_{ijl}(t+1)=1,\nonumber \\
w_{ijl}(t+1)=1.
\eea
Figure \ref{fig:SCgrowth} is a diagram illustrating one possible way the simplicial complex could grow in its first three time-steps. In this example we start with a single triangle at $t=1$. At $t=2$ a new node labelled $4$ is added and forms a triangle with the link $(1,2)$. At time $t=3$ no new nodes are added, but instead node $4$ gains an additional triangle formed with the link $(1,3)$.

\subsection{Number of links as a function of the number of nodes}
\begin{figure} 
\includegraphics[width=0.7\columnwidth]{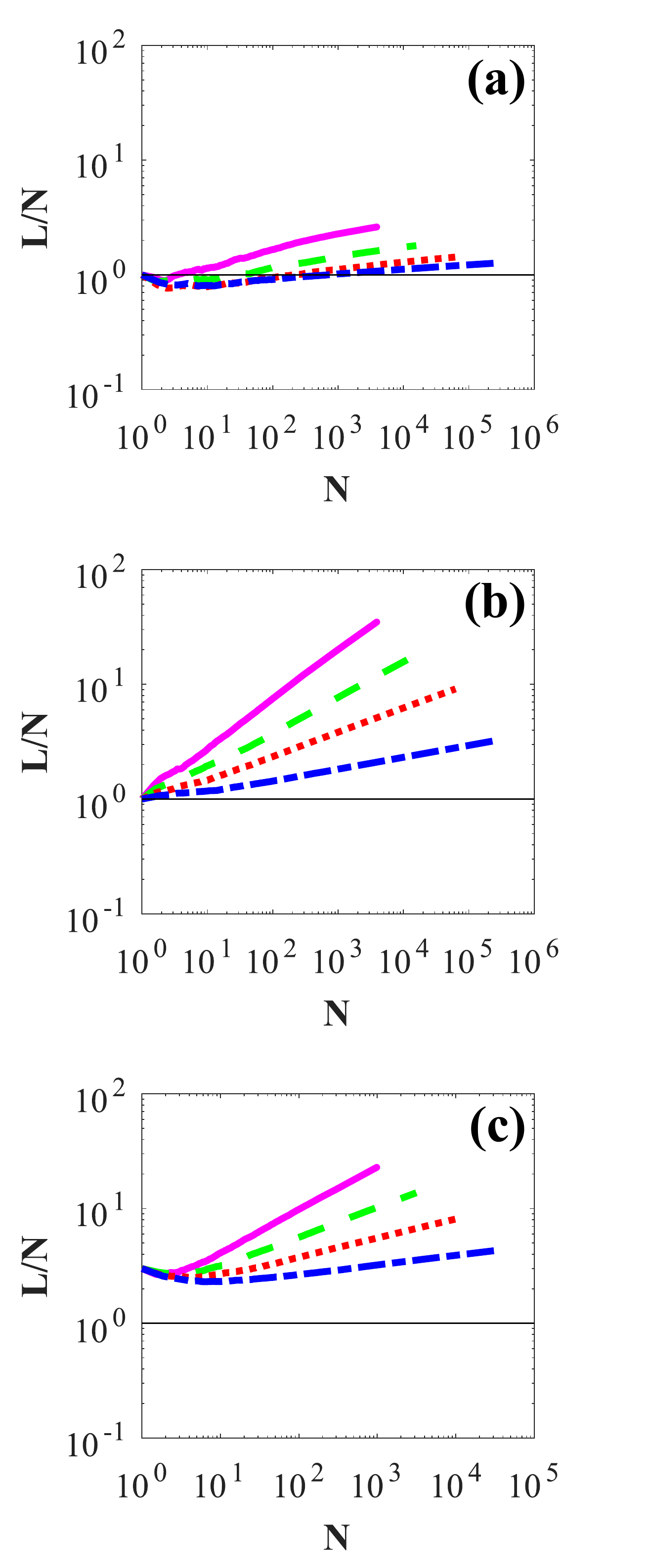}
\caption{(Color online). Total number of links over the total number of nodes, as a function of the total number of nodes, taken from simulation data. Panels (a), (b) and (c) show the results for the undirected network, directed network and simplicial complex respectively. For each model version and each choice of parameter $\alpha$, 50 realizations of the network were generated and averaged over. The results were obtained for $t=10^6$ and for $\alpha=0.6$, (purple solid line), $\alpha=0.7$ (green dashed line), $\alpha=0.8$ (red dotted line), and $\alpha=0.9$ (blue dot-dashed line).}
\label{fig:LvsN} 
\end{figure}

In all three versions of the model, the distributions of the strengths (or out-strengths) of the nodes are generated by a Pitman-Yor process and so therefore have `dense' power-law exponents $\delta \in (1,2)$. However, as mentioned before, this does not guarantee that the networks themselves are dense as the links can be weighted multiple times.
Therefore we have run extensive simulations of the three versions of the model  to investigate whether the total number of links grows super-linearly with the number of nodes. Figure \ref{fig:LvsN} shows how the number of links grows with the total number of nodes for a range of values of parameter $\alpha$ and for the three different versions of the model. We see that for all three versions the total number of links grows faster than the total number of nodes, indicating that the model produces dense networks and simplicial complexes. Moreover, as expected, the directed version of the model allows for the exploration of cases in which the ratio between the number of links and the number of nodes increases more rapidly than for the undirected version of the model.

\section{Strengths of the nodes}\label{sec:strengths}
The mean-field approximation is known to give very reliable results in the context of sparse growing network models. Therefore it is natural to approach the study of dense growing networks and simplicial complexes with the same techniques. Specifically here our goal is to derive the distribution of the strength $s$ in the undirected network model, the distribution of the out-strength $s^{out}$ in the directed network model and the distribution of the generalized out-strength $\tilde{s}^{out}$ for the simplicial complex model using the mean-field approximation.

\subsection{Evolution of the number of nodes and of the strengths}

The mean-field differential equations for the three cases differ only trivially, and therefore we will treat them using a unified set of equations that apply to all three cases. To this end we use the symbol $\hat{s}_i$ that indicates $s_i,s_i^{out}, \tilde{s}_i^{out} $ for the undirected network, directed network and directed simplicial complex versions of the model respectively.
Similarly the  Pitman-Yor probabilities $\Pi^{U}_i, \Pi_i^{D}$ and $\Pi_i^{SC}$ can be unified in a single expression 
\bea
\Pi_i = \left\{ \begin{array}{lll} \frac{\hat{s}_i(t)-\alpha}{(2-a)t} & \mbox{for} &1\leq i\leq N \\
 \frac{\alpha (N-a-b) }{(2-a)t} & \mbox{for} & i=N+1.
\end{array}\right. \label{PiUnified}
\eea
where we have introduced the parameters $a$ and $b$ taking values: $a=b=0$ for the undirected network case; $a=1,b=0$ for the directed network case and   $a=1,b=1$  for the directed simplicial complex case. Eq. (\ref{PiUnified}) thus subsumes Eq.s (\ref{PiU}), (\ref{PiD}) and (\ref{PiSC}) for the probabilities of a node $i$ being selected (or created) at time $t+1$ in the undirected, directed and simplicial complex versions of the model respectively.

As usual in the mean-field approximation, we will treat our stochastic variables $N(t)$, $\hat{s}_i(t)$ as deterministic continuous variables equal to their expected value over different realizations of the network or simplicial complex evolution. Since at each time the number of nodes to be chosen according to the Pitman-Yor probability is $(2-a)$ the mean-field equation determining the growth of the number of nodes in the network is given by 
\bea
\frac{dN}{dt}=(2-a)\Pi_{N+1} =\frac{\alpha (N-a-b)}{t},
\label{Ndiffeqn}
\eea
 with the initial condition
\bea
N(t=1)=2 +b.
\eea
The solution is then
\bea
N(t)=(2-a)t^{\alpha}+a+b.
\label{NMF}
\eea
The differential equations for $\hat{s}_i$ of a node $i$ born at time $t_i>1$ is given by 
\bea
\frac{d\hat{s}_i}{dt}=(2-a)\Pi_i=\frac{\hat{s}_i-\alpha}{t},
\label{Sdiffeqn}
\eea
with initial condition
\bea
\hat{s}_i(t_i)=1.
\eea
Therefore $\hat{s}_i$ increases linearly with time, and is given by 
\bea
\hat{s}_i=(1-\alpha)\left(\frac{t}{t_i}\right)+\alpha.
\label{SMF}
\eea

\subsection{Strength distribution}
The strength distribution can be easily derived in the mean-field approximation by using the mean-field expressions for the number of nodes $N(t)$ (Eq. $\ref{NMF}$) and the strength $\hat{s}(t)$ (Eq. $(\ref{SMF})$) as a function of time $t$.

\begin{figure} 
\includegraphics[width=0.7\columnwidth]{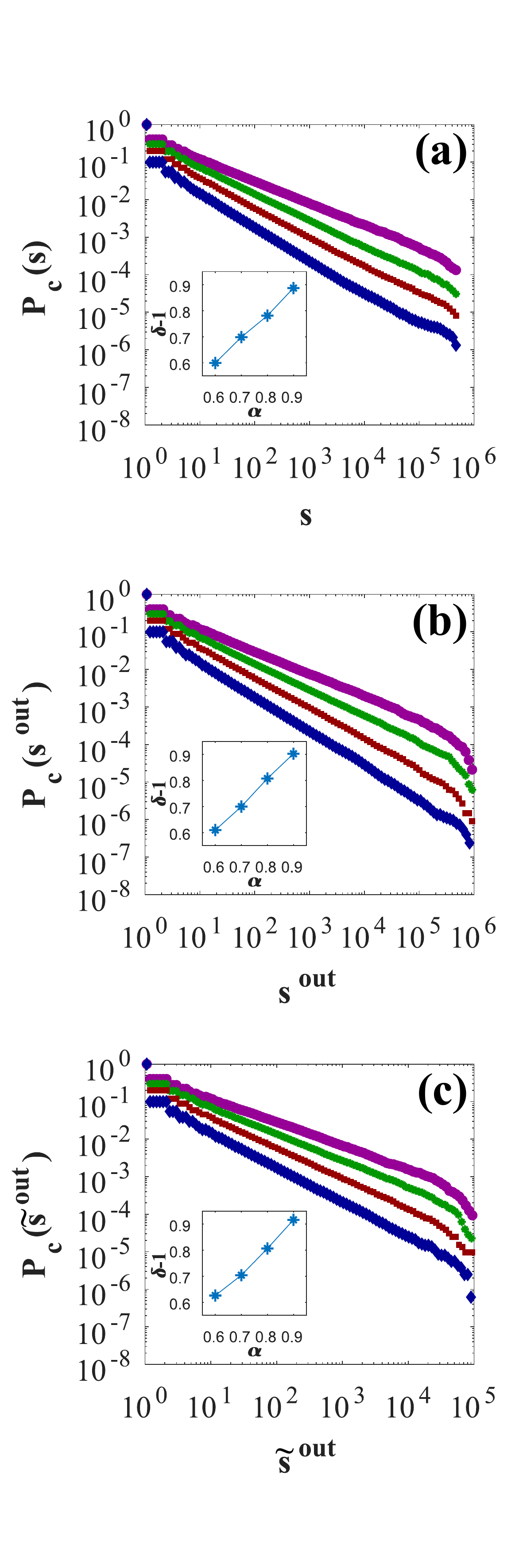}
\caption{ (Color online). Cumulative strength, out-strength and generalized out-strength distributions for the three versions of the model. For each version of the model and for each choice of parameter $\alpha$, 50 realizations of the network were generated and averaged over. The simulated results were obtained for $t=10^6$ and are represented by purple circles ($\alpha=0.6$), green stars ($\alpha=0.7$), red squares ($\alpha=0.8$) and blue diamonds ($\alpha=0.9$). The insets show the fitted exponents of the distributions for the four values of $\alpha$.}
\label{fig:Sdist} 
\end{figure}

To this end by using Eq. (\ref{SMF}) we first note that the  cumulative strength distribution $P(\hat{s}_i>\hat{s})$ indicating the probability that a random node $i$ has a strength $\hat{s}_i(t)>\hat{s}$ can be written as 
\bea
P(\hat{s}_i\geq\hat{s})=P\left(t_i\leq t^{\star}(\hat{s})\right),
\label{cumulativeS}
\eea
where $P\left(t_i\leq t^{\star}(\hat{s})\right)$ is the probability that a random node $i$ arrives in the network at time $t_i\leq t^{\star}(\hat{s})$ and where 
$t^{\star}(\hat{s})$ satisfies
\bea
\hat{s}=(1-\alpha)\left(\frac{t}{t^{\star}}\right)+\alpha.
\eea
Moreover we observe that the probability $P\left(t_i\leq t^{\star}(\hat{s})\right)$ is simply given by the fraction of nodes arrived in the network before time $ t^{\star}(\hat{s})$, i.e.
\bea
P\left(t_i\leq t^{\star}(\hat{s})\right)= \frac{N(t^{\star}(\hat{s}))-a-b}{N(t)-a-b}.
\eea
The strength distribution $\tilde{P}(\hat{s})$ is thus given by
\bea
\tilde{P}(\hat{s})&=&\frac{d}{d\hat{s}} \left[ 1 - P(\hat{s}_i\geq \hat{s})\right]\nonumber \\
&\simeq &\frac{\alpha}{1-\alpha}\left(\frac{1-\alpha}{\hat{s}}\right)^{\alpha+1},
\label{PSMF}
\eea
where the last expression is valid for $\hat{s}\gg 1$.
Therefore the strength distribution is power-law distributed with exponent $1+\alpha\in (1,2]$. Figure \ref{fig:Sdist} shows the strength distributions arising from simulations of the three models. We see that in all three versions of the model, and for all values of $\alpha$ used, that the strength distributions follow a power-law. In the insets of each panel we see that the exponents fitted to the distributions are very close to $1+\alpha$ as predicted by Eq. (\ref{PSMF}).

\section{Reinforcement probabilities}\label{sec:reinforce}
By reinforcement probability we mean the probability that at time $t$ we either add a new link (or a new triangle) or in the case where the link (or triangle) already exists that we increase its weight.
This probability can be directly calculated in the mean-field approximation using Eqs. (\ref{NMF}) and (\ref{SMF}).\\

In the undirected case  we indicate with $\pi_{ij}(t,t_i,t_j)$ the probability that at time $t$ we reinforce or add a link between node $i$ and node $j$ given that node $i$ and node $j$ have been added to the network at time $t_i$ and $t_j$ respectively.
The reinforcement probability is therefore given by 
\bea
\pi_{ij}^{U}(t,t_i,t_j)=2\Pi^{U}_i\Pi^{U}_j\eea
where $\Pi_i^{U}, \Pi_j^{U}$ are calculated at time $t$. Therefore we have
\bea
\pi_{ij}^{U}(t,t_i,t_j)=\frac{\left(s_i (t) - \alpha\right)\left(s_j (t) - \alpha \right)}{2t^2},
\label{PIUpreMF}
\eea
which by inserting the mean-field expression for $s_i(t)$ gives
\bea
\pi_{ij}^{U}(t,t_i,t_j)=\frac{\left(1 - \alpha\right)^2}{2 t_i t_j}.
\label{PIUMF}
\eea
Similarly it can be shown that in the directed case the probability $\pi_{ij}^{D}(t,t_i,t_j)$ that a link from node $i$ to node $j$ is reinforced at time $t$ given that nodes $i$ iand $j$ are arrived in the network at time $t_i$ and $t_j$ respectively can be expressed as
\bea
\pi_{ij}^{D}(t,t_i,t_j)=\frac{1}{N(t)}\frac{{s}_i^{out} (t) - \alpha}{t},
\label{PIDpreMF}
\eea
which using the mean-field solution of ${s}_i^{out}$ and $N(t)$ gives
\bea
\pi_{ij}^{D}(t,t_i,t_j)=\frac{1-\alpha}{(t^\alpha -1) t_i}.
\label{PIDMF}
\eea
For the simplicial complex we indicate with $\tilde{\pi}_{i,\ell}(t,t_i,\tau_{\ell})$ the probability that a triangle with source node $i$ and target link $\ell$ is reinforced, conditioned on their respective birth times $t_i$ and $\tau_{\ell}$. We write this as
\bea
\tilde{\pi}_{i,\ell}(t,t_i,\tau_{\ell})&=&\frac{\tilde{s}_i^{out}-\alpha}{t}\frac{1}{L(t)},
\label{PISCpreMF}
\eea
where $L(t)$ is the total number of (directed) links at time $t$. Using Eq. (\ref{SMF}) we obtain
\bea
\tilde{\pi}_{i,\ell}(t,t_i,\tau_{\ell})&=&\frac{1-\alpha}{t_i}\frac{1}{L(t)}.
\label{PISCMF}
\eea

\section{Degree distribution}\label{sec:degrees}
In this section we use the mean-field results of sections \ref{sec:strengths} and \ref{sec:reinforce} to derive equations for the degrees of the nodes conditioned on their birth times. We evaluate these equations numerically for a range of values of the parameter $\alpha$ and find power-law scalings with $\gamma=2$ for $k_i$ in the undirected case, and $\gamma<2$ for $k_i^{out}$ and $\tilde{k}_i^{out}$ in the directed and simplicial complex cases. We also compare our numerically obtained predictions with simulation results, validating our mean-field approach. 

We observe that for the studied model the mean-field solution provides a very good approximation to the simulation results. Although we do not have a quantitative theoretical argument to show the efficacy of the mean-field approximation the results is not completely surprising. In fact already for the BA model and the Simon model the mean-field approximation works very well due to the presence of the reinforcement dynamics. In the Pitman-Yor process and in the proposed dense network model the balance between the reinforcement move and the ``innovation" move (placing a new ball in a new box or attaching a new link to a new node) is further shifted in favor of the reinforcement dynamics. This feature therefore could potentially justify the efficacy of the mean-field approximation.

\subsection{Undirected network case}
\begin{figure} 
\includegraphics[width=1\columnwidth]{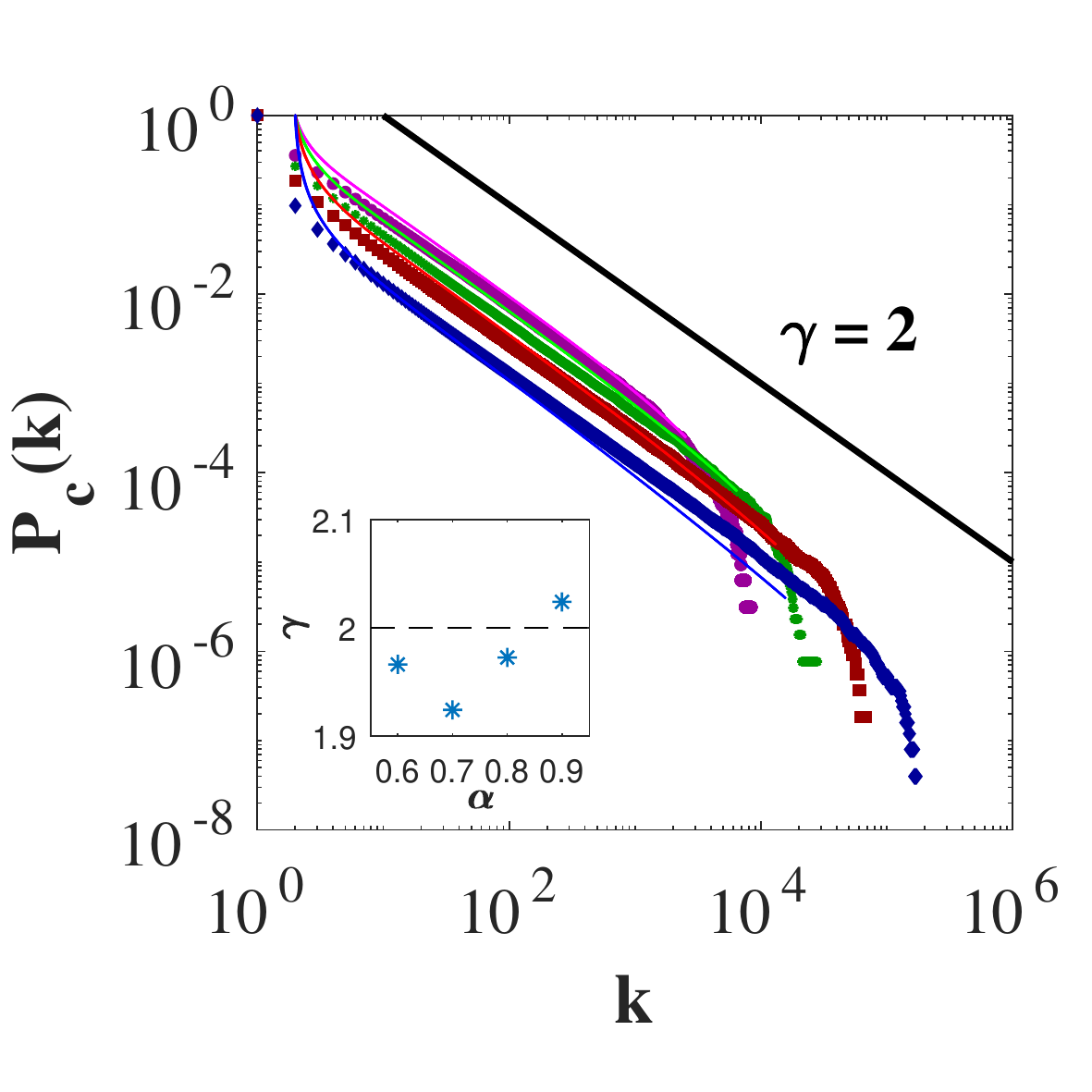}
\caption{ (Color online). Cumulative degree distributions in the undirected case. For each choice of parameter $\alpha$, 50 realizations of the network were generated and averaged over. The simulated results were obtained for $t=10^6$ and are represented by purple circles ($\alpha=0.6$), green stars ($\alpha=0.7$), red squares ($\alpha=0.8$) and blue diamonds ($\alpha=0.9$). The numerical results are represented by the purple solid line ($\alpha=0.6$), green dashed line ($\alpha=0.7$), red dotted line ($\alpha=0.8$) and blue dot-dashed line ($\alpha=0.9$). The solid black line shows an exact power-law with exponent $\gamma=2$ for comparison. The inset shows the fitted exponents of the simulated distributions for the four values of $\alpha$.}
\label{fig:undirected_deg_dist} 
\end{figure}

We write the degree of a node $i$ born at time $t_i$ in terms of the link probabilities $p_{ij}(t,t_i,t_j)$:
\bea
k_i (t,t_i)&=& \int_{1}^{t} dt_j \dot{N}(t_j)p_{ij}(t,t_i, t_j).
\label{KUMF}
\eea
The link probability $p_{ij}(t,t_i,t_j)$ is the probability that a link exists between nodes $i$ and $j$ conditioned on their birth times $t_i$ and $t_j$, and may be written as
\bea
p_{ij}(t,t_i,t_j) &=&1- \prod_{t'=\tau}^{t}\left[1-\pi_{ij}(t',t_i,t_j)\right],
\label{linkprobA}
\eea
where $\pi_{ij}(t',t_i,t_j) $ is the reinforcement probability given in (\ref{PIUMF}) and $\tau=\max\{t_i,t_j\}$ is the first time that both $i$ and $j$ are present in the network, i.e. (\ref{linkprobA}) is $1$ minus the probability that the pair $(i,j)$ is not reinforced in the time interval $[\tau,t]$. The $\pi_{ij}$ reinforcement probabilities are very small for almost all pairs of nodes, so we make the approximation
\bea
\prod_{t'=\tau}^{t}\left[1-\pi_{ij}(t',t_i,t_j)\right] \simeq \exp\left(-\sum_{t'=\tau}^{t} \ \pi_{ij}(t',t_i,t_j) \right).
\eea
Taking $t$ to be very large, we approximate the sum with an integral, and write (\ref{linkprobA}) as
\bea
p_{ij}(t,t_i,t_j) \simeq 1-\exp\left(-\int_{\tau}^{t}dt' \ \pi_{ij}(t',t_i,t_j) \right).
\label{linkprob}
\eea
It is then straight-forward to obtain the following expression for the degree from Eq.s (\ref{linkprob}) and (\ref{PIUMF}):
\bea
k_i(t,t_i) & \simeq & \alpha\int_1^{t_i} dt_j t_j^{\alpha-1} \left[ 1- e^{-\frac{(1-\alpha)^2}{2 t_i t_j}(t - t_i)} \right] \nonumber
\\ & & + \alpha \int_{t_i}^{t} dt_j t_j^{\alpha-1} \left[ 1- e^{-\frac{(1-\alpha)^2}{2 t_i t_j}(t - t_j)} \right].
\label{KUMF2}
\eea
 Eq. (\ref{KUMF2}) may be written as
\bea
k_i(t,t_i) & \simeq & \alpha A^\alpha (t-t_i)^\alpha \int_{A (\frac{t}{t_i}-1)}^{A(t-t_i)} dx x^{-(1+\alpha)} \left[ 1- e^{-x} \right]
\\ & & + \alpha A^\alpha t^\alpha \int_{0}^{A (\frac{t}{t_i}-1)} dx (A + x)^{-(1+\alpha)} \left[ 1- e^{-x} \right],
\label{KUMF3}
\eea
where $A=\frac{(1-\alpha)^2}{2 t_i}$ and in the first integral we have used the change in variable $x=A(t - t_i) t_j^{-1}$ while in the second integral we have used $x=A(\frac{t}{t_j} -1)$. Taking $B=(1-\alpha)^2/2\ll 1$ this equation can be approximated by the scaling function
\bea
k_i(t,t_i)  & \simeq &G(t,t/t_i)
\eea
with 
\bea
G(t,t/t_i=y)&=&\alpha B^\alpha (y-1)^\alpha \int_{By(y-1)/t}^{B(y-1)} dx x^{-(1+\alpha)}  \left[ 1- e^{-x} \right]
\nonumber \\ & & + \alpha B^\alpha y^\alpha \int_{0}^{By (y-1)/t} dx x^{-(1+\alpha)} \left[ 1- e^{-x} \right].\nonumber
\label{KUMF3}
\eea
Finally by taking the limit $t,t_i \to \infty $ with $t/t_i$ fixed, it is possible to show  that $k_i(t,t_i)$ becomes exclusively a function of $t/t_i$
\bea
k_{i}(t,t_i)=\tilde{G}(t/t_i)
\eea
where for $t/t_i\gg1$ 
\bea
\tilde{G}(t/t_i) \propto \left(\frac{t}{t_i}\right)^{\alpha}.
\label{KUMF5}
\eea
Therefore the tail of the cumulative degree distribution $P(k_i>k)$ can be obtained within the mean-field approximation using 
\bea
P(k_i>k)=\frac{N\left(t^{\star}(k)\right)}{t^\alpha}=\left(\frac{t^{\star}(k)}{t}\right)^\alpha ,
\label{KUPK1}
\eea
where $t^{\star}(k)$ is the birth time such that 
\bea
k_i\big(t,t^{\star}(k)\big)=k .
\label{KUtstar}
\eea
From Eq.s (\ref{KUMF5}) and (\ref{KUtstar}) we obtain the scaling
\bea
P(k_i>k) \propto k^{-1},
\eea
indicating that for large $t$, the degree distribution $P(k)$ has a power-law tail with exponent $\gamma = 2$. We confirm these results by comparing them to simulations of the process for a range of values of $\alpha$. Figure \ref{fig:undirected_deg_dist} shows the average cumulative degree distributions given by the simulations. Also included are theoretical predictions obtained by evaluating Eq. (\ref{KUMF2}) numerically. The inset plot shows the values of $\gamma$ obtained from fitting power-laws to the tails of the simulation data. We see that for all the degrees of the nodes follow power-laws with values of $\gamma$ close to the theoretical prediction of $2$.

\begin{figure} 
\includegraphics[width=1\columnwidth]{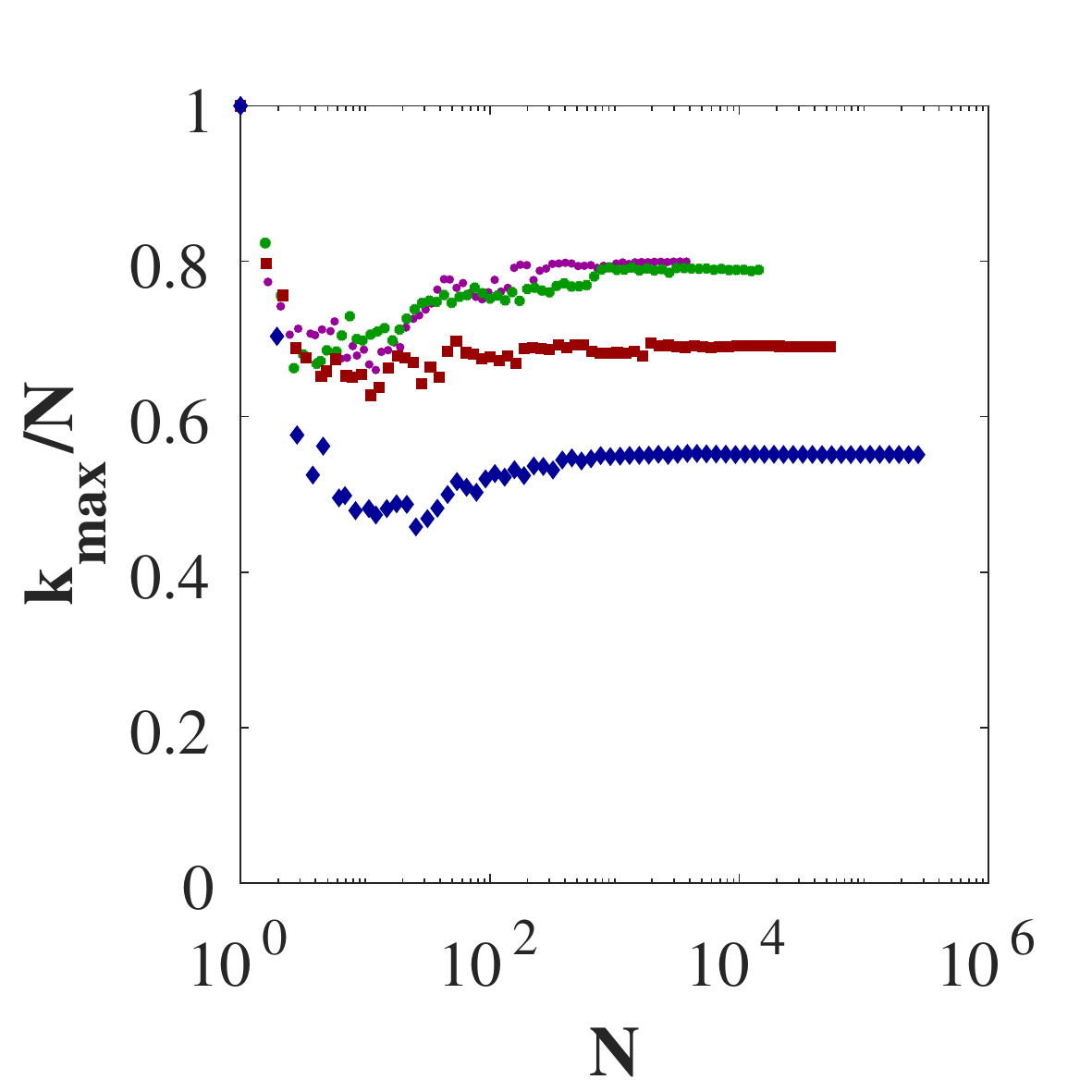}
\caption{ (Color online). Evolution of the average normalized degree cut-off for the undirected case as a function of average total number of nodes in the network. For each choice of parameter $\alpha$, 50 realizations of the network were generated and averaged over. The simulated results were obtained for $t=10^6$ and are represented by purple circles ($\alpha=0.6$), green stars ($\alpha=0.7$), red squares ($\alpha=0.8$) and blue diamonds ($\alpha=0.9$). }
\label{fig:cutoff} 
\end{figure}

Additionally  we have studied the scaling of the maximum degree (cutoff) $k_{max}$ as a function of the network size $N$.
This study reveals that the  cutoff scales with the network size $N$ with a proportionality constant depending on the value of $\alpha$ (see Figure $\ref{fig:cutoff}$).

\subsection{Directed network case}
\begin{figure} 
\includegraphics[width=1\columnwidth]{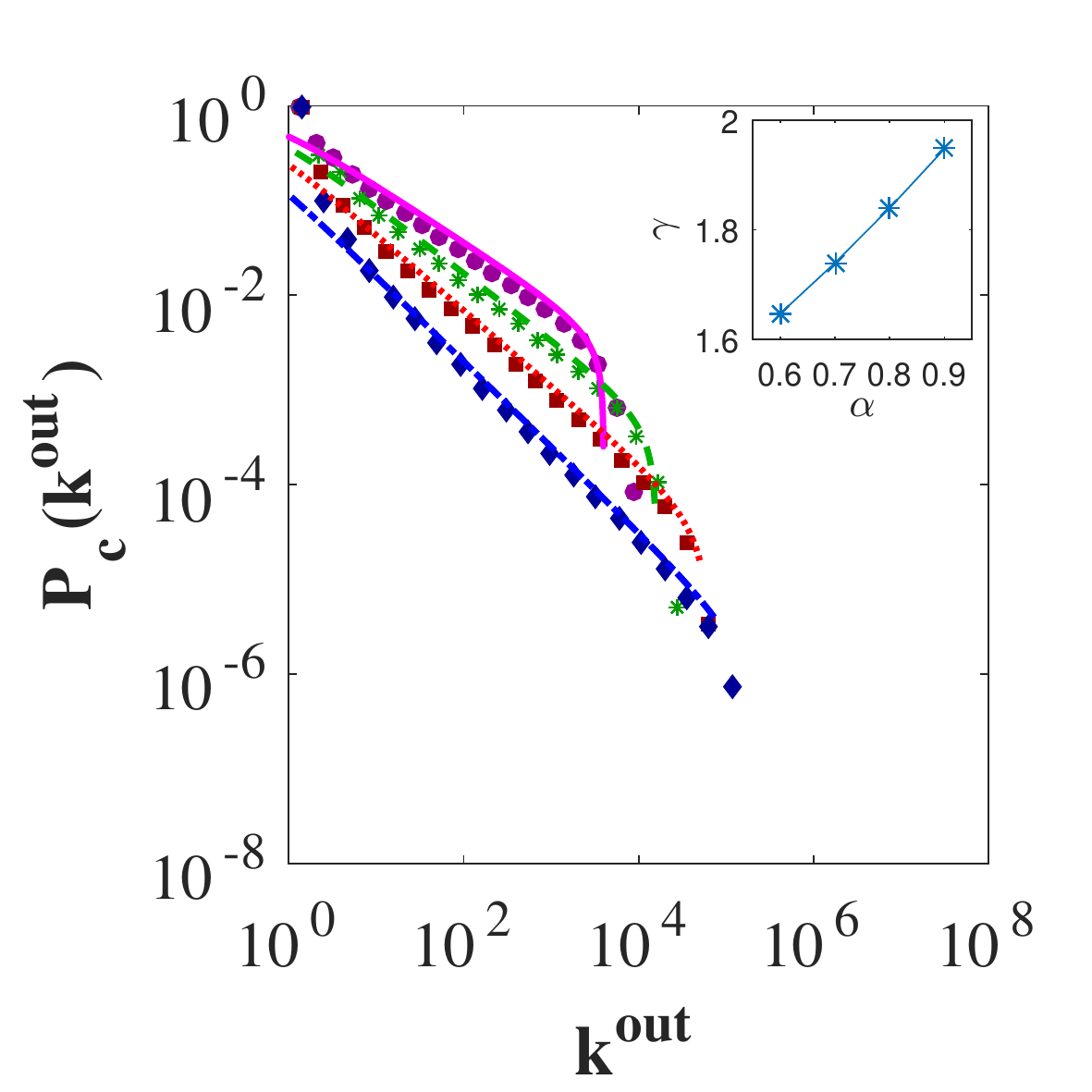}
\caption{{ {(Color online). Cumulative distributions of the out-degrees. For each choice of parameter $\alpha$, 50 realizations of the network were generated and averaged over. The simulated results were obtained for $t=10^6$ and are represented by purple circles ($\alpha=0.6$), green stars ($\alpha=0.7$), red squares ($\alpha=0.8$) and blue diamonds ($\alpha=0.9$). The numerical results are represented by the purple solid line ($\alpha=0.6$), green dashed line ($\alpha=0.7$), red dotted line ($\alpha=0.8$) and blue dot-dashed line ($\alpha=0.9$). The inset shows the fitted exponents of the simulated distributions for the four values of $\alpha$.}}} 
\label{fig:directed_deg_dist} 
\end{figure}

The out-degree $k^{out}_i(t,t_i)$ at time $t$ of node $i$ born at time $t_i$ in the directed case is
\bea
k^{out}_i(t,t_i)\simeq\int_1^{t} dt_j \dot{N}(t_j)p_{ij}(t,t_i,t_j),
\label{KDMF}
\eea
where $p_{ij}(t,t_i,t_j)$ is as in Eq. (\ref{linkprob}) but with $\pi_{ij}(t',t_i,t_j)$ given  by Eq. (\ref{PIDMF}). 
Let us use the direct evaluation of the integral
\bea
\int_{\tau}^{t}dt' \ \pi_{ij}(t',t_i,t_j)&=&\frac{1}{t_i}[t^{1-\alpha}-\tau^{1-\alpha}]\nonumber \\
&=&{t^{-\alpha}}\frac{t}{t_i}\left[1-\left(\frac{\tau}{t}\right)^{1-\alpha}\right]
\eea 
to express  the out-degree $k_i^{out} (t)$ of a node $i$ with arrival time $t_i$ at time $t$ (\ref{KDMF}) as
\bea
k_i^{out} (t)&=& N(t_i)p(t,t_i,t_i) + \int_{t_i}^{t} d\tau \frac{\alpha N(\tau) }{\tau }p(t,t_i,\tau).\nonumber \\
\eea
Moreover integrating by parts we find
\bea
\frac{k_i^{out}(t)}{t^{\alpha}}&=&1-\left(\frac{t_i}{t}\right)^{\alpha}e^{-\tilde{A}(t,t_i/t)\left[1-\left(\frac{t_i}{t}\right)^{1-\alpha}\right]}\nonumber \\
&&-\alpha\int_{t_i/t}^{1} dx  x^{\alpha-1}e^{-\tilde{A}(t,t/t_i)[1-x^{1-\alpha}]}\nonumber \\
&=&(1-\alpha)\tilde{A}(t,t_i/t) \int_{t_i/t}^{1} dx e^{-\tilde{A}(t,t/t_i)[1-x^{1-\alpha}]}
\label{KMFptau}
\eea
where 
\bea
\tilde{A}(t,t/t_i)=t^{-\alpha}\left(\frac{t}{t_i}\right).
\eea
Therefore we find that as $t,t_i\to \infty$ with $t/t_i$ fixed, 
\bea
k_i^{out}(t)&\simeq & (1-\alpha)\left[\frac{t}{t_i}-1\right]
\eea
Therefore the cumulative degree distribution may be found from 
\bea
P(k_i(t)>k)=P(t_i<t^{\star}(k)))
\eea
with 
\bea
t^{\star}(k)=t \frac{(1-\alpha)}{(k+1-\alpha)}
\eea
and
\bea
P(t_i<t^{\star}(k)))=\frac{N(t^{\star}(k))}{N(t)}.
\eea
Using the mean-field expression for the number of nodes given by Eq. (\ref{NMF}) we find the cumulative out-degree distribution to be
\bea
P(k_i>k)=\left(\frac{(1-\alpha)}{(k+1)}\right)^{\alpha},
\eea
which implies the out-degree distribution is
\bea
P(k)=\alpha\frac{(1-\alpha)^{\alpha}}{(k+1)^{\alpha+1}}\simeq k^{-\alpha-1}.
\eea
Therefore, we see that within our mean-field approximation the distribution of out-degrees has a power-law tail with exponent $\gamma=1+\alpha$.
Figure \ref{fig:directed_deg_dist} shows theoretical predictions for the full cumulative out-degree distribution, obtained from numerical evaluation of Eq. (\ref{KDMF}) for a selection of values of $\alpha$. Also included in the figure are the results of simulations for the same values of $\alpha$. The inset plot shows the values of $\gamma$ obtained from fitting power-laws to the tails of the simulation data. We see that the out-degrees of the nodes follow power-laws with increasing values of $\gamma$ for larger $\alpha$, and with all values of $\gamma$ between $\gamma=1$ and $\gamma=2$. From the inset plot it is clear that the exponents of tails of the distributions closely agree with the theoretical prediction of $\gamma=1+\alpha$.

\subsection{Directed simplicial complex case}
In the case of the simplicial complex, the generalized out-degree $\tilde{k}^{out}_i(t,t_i)$ of a node $i$ is the number of triangles for which $i$ is the source node. In the mean-field approximation we may write this as
\bea
\tilde{k}_i^{out}(t,t_i)\simeq\int_1^{t} d\tau_{\ell} \dot{L}(\tau_{\ell}) \hat{p}(t,t_i,\tau_{\ell})
\label{KSintro}
\eea
where 
\bea
\hat{p}(t,t_i,\tau_{\ell})=1-e^{-(1-\alpha)/t_i\int_{\max(t_i,\tau_{\ell})}^{t}dt' [L(t')]^{-1}} \label{pSC}
\eea
is the probability of a triangle with source node $i$ and target link $l$. From figure \ref{fig:LvsN} its clear that for large $t$, $L(t)$ grows like a power of $t$. We therefore assume 
\bea
L(t) = c t^b \label{LPL}
\eea
and obtain values $c$ and $b$ for each choice of $\alpha$ by fitting Eq. (\ref{LPL}) to the data shown in figure \ref{fig:LvsN}(c). Substituting Eq. (\ref{LPL}) in to (\ref{pSC}) we obtain the following for the generalized out-degree of a node $i$ born at time $t_i$,

\bea 
\begin{array}{lll} {\tilde{k}}_i^{out}(t,t_i)= & & 
\\ c b \int_{t_i}^{t} d\tau_{\ell} \tau_{\ell}^{b-1}\left[1 - e^{-\frac{1-\alpha}{c(1-b)t_i}\left(t^{1-b} - \max(t_i,\tau_{\ell})^{1-b}  \right)}\right]. & &
\end{array}\label{KSC}
\eea

\begin{figure} 
\includegraphics[width=1\columnwidth]{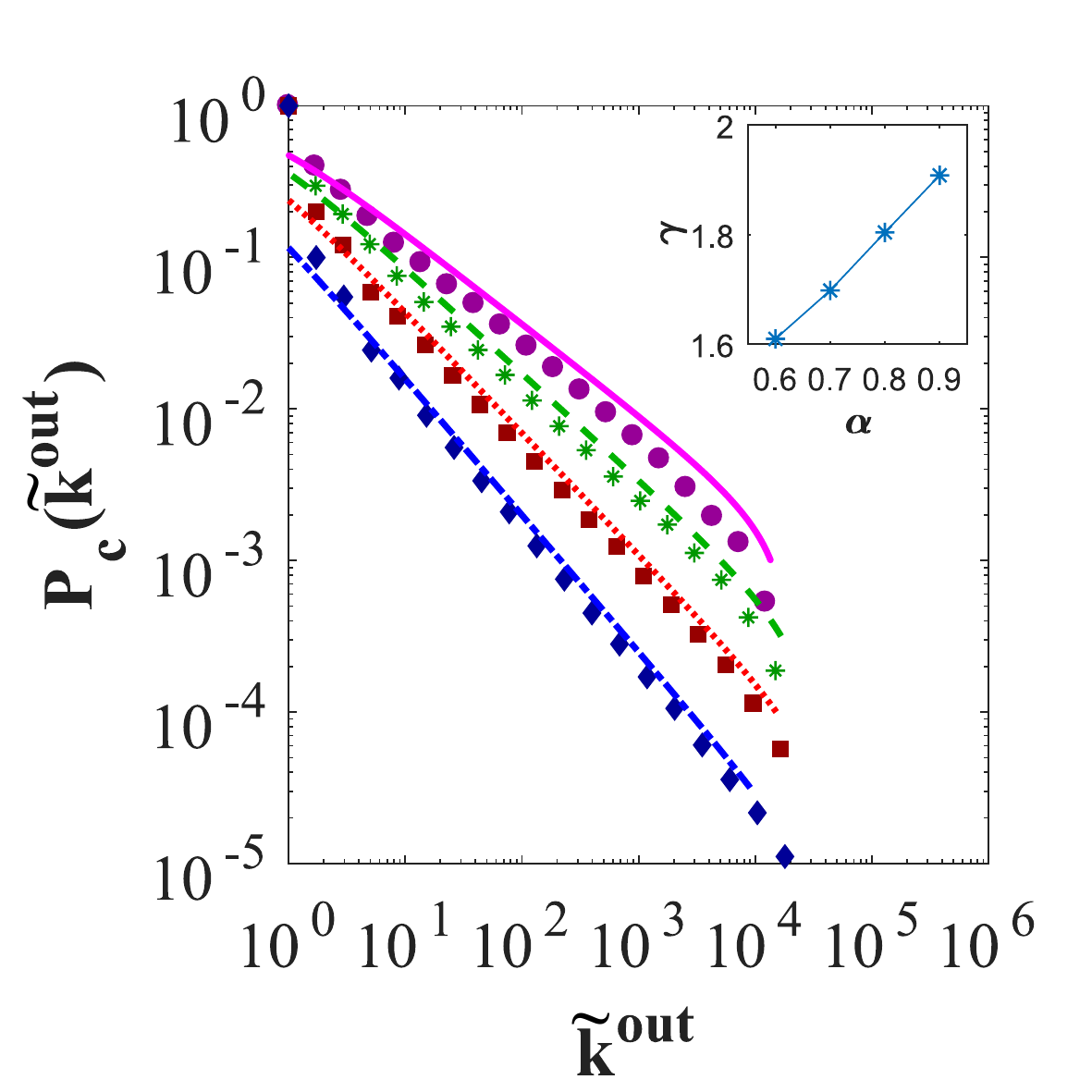}
\caption{{ {(Color online). Cumulative distributions of the generalized out-degrees. For each choice of parameter $\alpha$, 50 realizations of the simplicial complex were generated and averaged over.  The simulated results were obtained for $t=10^5$ and are represented by purple circles ($\alpha=0.6$), green stars ($\alpha=0.7$), red squares ($\alpha=0.8$) and blue diamonds ($\alpha=0.9$). The numerical results are represented by the purple solid line ($\alpha=0.6$), green dashed line ($\alpha=0.7$), red dotted line ($\alpha=0.8$) and blue dot-dashed line ($\alpha=0.9$). The inset shows the fitted exponents of the simulated distributions for the four values of $\alpha$.}}} 
\label{fig:SC_deg_dist} 
\end{figure}

Figure \ref{fig:SC_deg_dist} shows theoretical predictions for the full cumulative generalized out-degree distribution, obtained from numerical evaluation of Eq. (\ref{KSC}) for a selection of values of $\alpha$. Also included in the figure are the results of simulations for the same values of $\alpha$. The inset plot shows the values of $\gamma$ obtained from fitting power-laws to the tails of the simulation data.
We see that the generalized out-degrees of the nodes follow power-laws with increasing values of $\gamma$ for larger $\alpha$, and with all values of $\gamma$ between $\gamma=1$ and $\gamma=2$. From the inset plot we see that the exponents of tails of the distributions are quite close to the exponents of the generalized out-strengths $\delta=1+\alpha$.

\section{Strength versus degree}\label{sec:SvsK}
\begin{figure} 
\includegraphics[width=0.7\columnwidth]{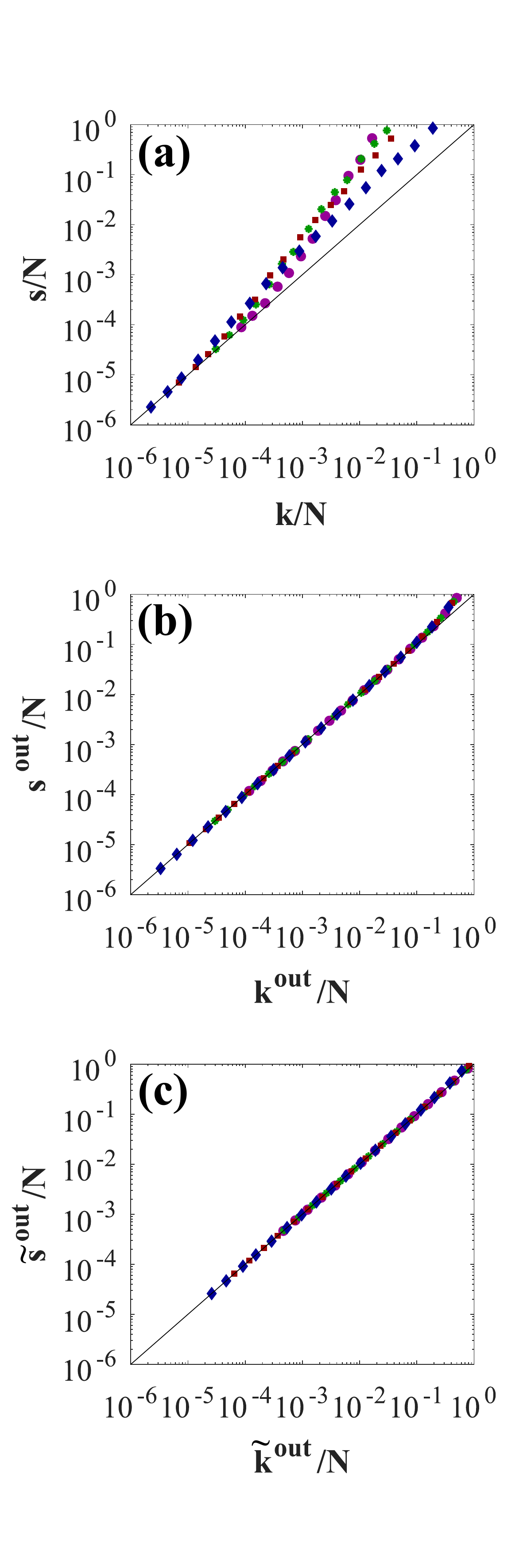}
\caption{{ {(Color online). Plot of the normalized strengths versus the normalized degrees. Panels (a), (b) and (c) show the results for the undirected, directed and simplicial complex versions respectively. For each choice of parameter $\alpha$, 50 realizations of the simplicial complex were generated and averaged over.  The results were obtained for $t=10^5$ and are represented by purple circles ($\alpha=0.6$), green stars ($\alpha=0.7$), red squares ($\alpha=0.8$) and blue diamonds ($\alpha=0.9$). The solid black line is the function $f(\frac{k}{N})= \frac{k}{N}$, and is there as a guide to see how closely the strengths match the degrees.}}}
\label{fig:SvsK} 
\end{figure}

The models we have introduced all produce networks or simplicial complexes where the strengths of the nodes and total number of nodes follow the statistics of the Pitman-Yor model of balls in boxes. In contrast, the degree statistics do not follow the Pitman-Yor model, as links (or triangles) between nodes may be weighted multiple times without altering the degrees (or generalized degrees) of the nodes. In section \ref{sec:degrees} we saw that in the directed network and directed simplicial complex versions, the exponents of the degree and generalized degree distributions are a fairly close match for the exponents of the strength distributions, while in the undirected version the exponent of the degree distribution is always equal to $2$. An interesting question is therefore what is the relation between the degrees and generalized degrees of the nodes and their strengths and generalized strengths? In particular, in the directed network and directed simplicial complex versions we would like to know to what extent the strengths and generalized strengths can act as proxies for the degrees and generalized degrees or whether the stregth increases super-linearly with the degree of the nodes \cite{Barrat,Owen2}. To this end we have run simulations of the models for various values of $\alpha$ with the aim of extracting the average relations over all of the realizations. The total number of nodes in a realization is a random variable and is in fact a non-self-averaging quantity \cite{SMCRP}. This means that over a set of realizations the total number of nodes can vary widely. This is important as the probability that a source node either gains a new link (or triangle) or has one of its existing ones reinforced depends on the ratio of the relative size of the degree of the node with respect to $N$. Therefore, to see the effect of this `crowding' of the links of high degree nodes we have normalized the strength and degree data by dividing by $N$ for each realization before averaging over all realizations. Figure \ref{fig:SvsK} shows the relation between the normalized strengths and degrees of the nodes for the three versions of the model. We see from panel a) that the strengths of the nodes in the undirected version are significantly higher than their degrees, with the effect being greater for nodes with higher degree. This is expected, as the probability of a link being reinforced more than once is larger when the strengths of it's two nodes are larger. In contrast, we see from panels b) and c) that the average out-strengths and average generalized out-strengths are very close to equal to the out-degrees and generalized out-degrees of the nodes in the directed network and simplicial complex versions respectively, suggesting the strengths may indeed act as proxies for the degrees.

\section{Clustering and Degree Correlations}\label{sec:C&A}
\begin{figure} 
\includegraphics[width=0.75\columnwidth]{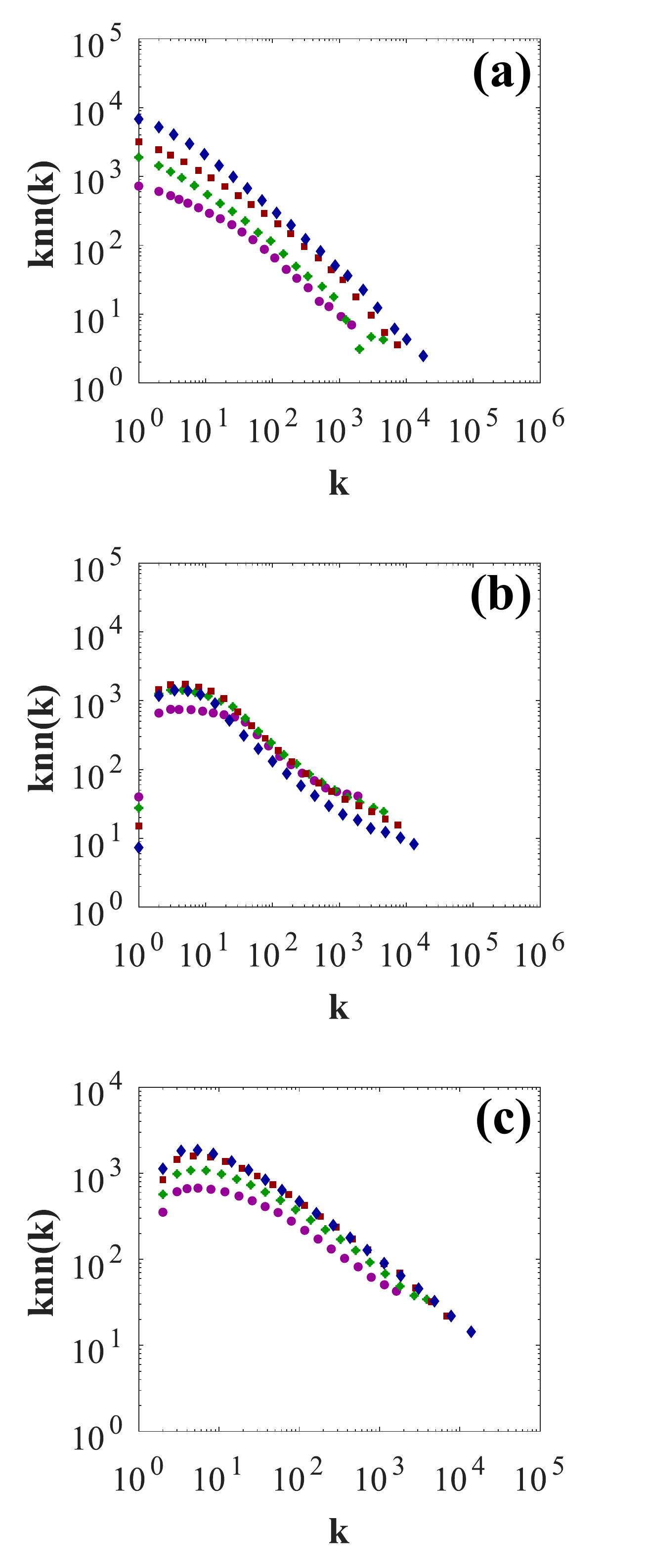}
\caption{{{(Color online). Average degree of the neighbours of nodes of degree $k$, taken from simulation data. Panels (a), (b) and (c) show the results for the undirected network, directed network and simplicial complex respectively. For each model version and each choice of parameter $\alpha$, 50 realizations of the network were generated and averaged over. The results were obtained for $t=10^5$ and for $\alpha=0.6$, (purple circles), $\alpha=0.7$ (green stars), $\alpha=0.8$ (red squares), and $\alpha=0.9$ (blue diamonds).}}} 
\label{fig:knn} 
\end{figure}

\begin{figure} 
\includegraphics[width=1\columnwidth]{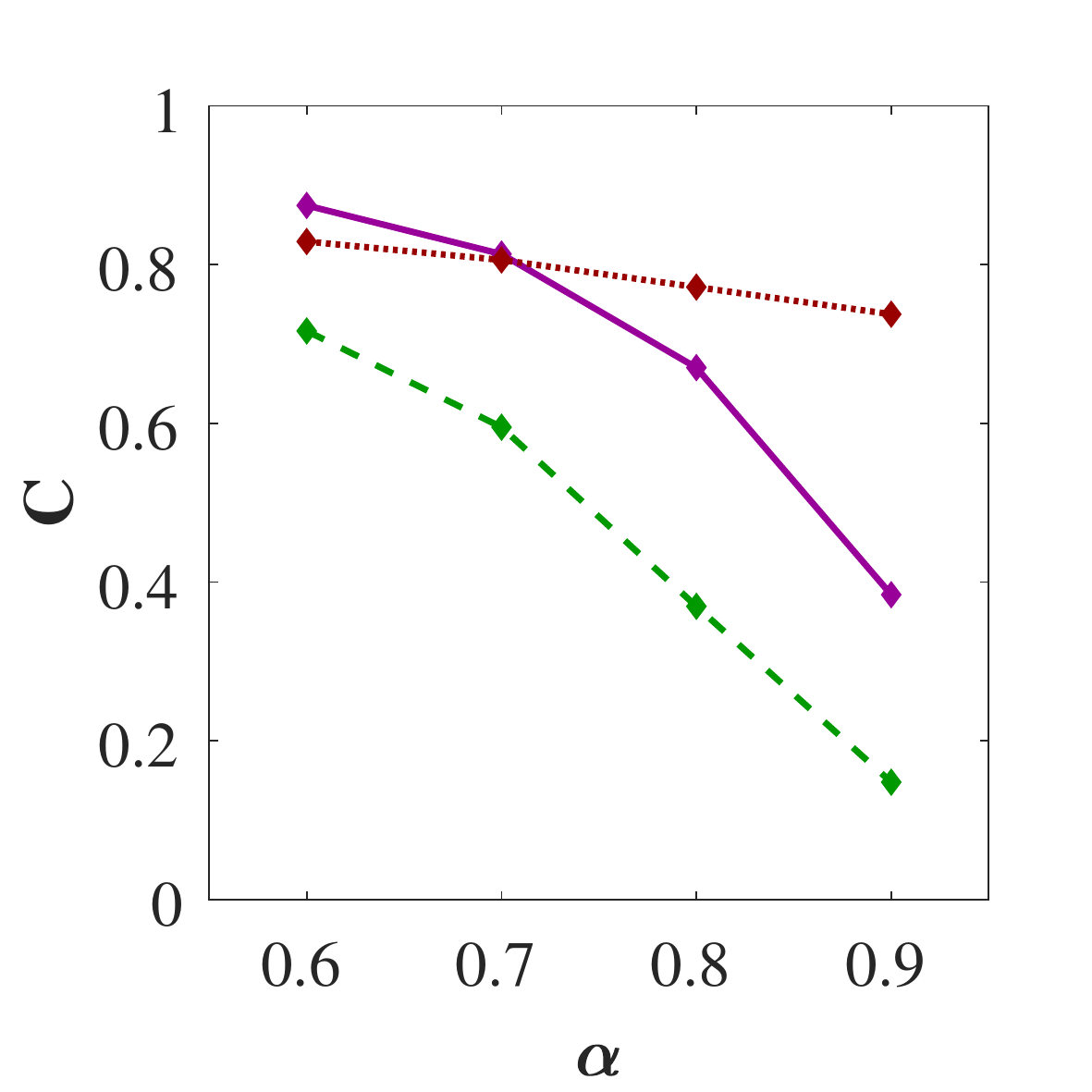}
\caption{{{(Color online). Average clustering of the nodes at different values of $\alpha$, taken from simulation data with $t=10^5$. For each model version and each choice of parameter $\alpha$, 50 realizations of the network were generated and averaged over. Results from the three model versions are represented by the purple solid line (undirected), green dashed line (directed) and red dotted line (simplicial complex).}}} 
\label{fig:clavg} 
\end{figure}

In this section we explore using simulations the clustering and degree correlations of the undirected (and unweighted) networks produced by the three versions of our model. In order to compare the results for the undirected network model with the results of  the directed network and directed simplicial complex versions, we decided to discard the information about the direction of the links in the directed network and directed simplicial complex versions. Figure \ref{fig:knn} shows the average degree $knn(k)$ of the neighbours of nodes with given degree $k$ for the three versions. We see that for all three versions of the model, the networks produced are strongly disassortative. Interestingly in the undirected version, despite the fact that in Sec. \ref{sec:degrees} we found that the degree distribution has the same power-law exponent for different values of $\alpha$, the strength of the disassortativity appears to be greater for increasing values of $\alpha$. A likely explanation for this trend is that for larger $\alpha$ there is bias away from adding links between existing high degree nodes and towards creating links between a new node and a high degree node.
Figure \ref{fig:clavg} shows the average clustering of all nodes in the networks against the model parameter $\alpha$ for the three versions of the model.  We see that the average clustering decreases with increasing values of $\alpha$ for all three versions.

\section{Conclusions}\label{sec:conclusions}
In this paper we have presented three similar models for producing dense networks with power-law distributions of the strengths and degrees. The growth mechanisms of these models are analogous to the Pitman-Yor process, a stochastic process well-known among probability theorists for generating random partitions with power-law distributions of block sizes. Our undirected model can in one sense be thought of as a network with multiedges and a power-law degree distribution with tunable dense exponent $\gamma= 1+\alpha \in (1,2)$ or in a different sense as a weighted network with a power-law degree distribution with the border-line dense exponent $\gamma = 2$. Our directed network model produces dense directed networks with out-degree distributions that follow a power-law with tunable exponent $\gamma= 1+\alpha$, and homogeneous distributions of the in-degrees. Our simplicial complex version extends the concept of a scale-free network to a scale-free simplicial complex with power-law distribution of the generalized out-degrees. These models demonstrate the difficulty in producing networks that are both dense and scale-free. However, they give insight into the possible mechanisms by which real-world networks densify, and may have a use as null-models for the growth of on-line social networks, recommendation networks or the brain. We show that the models are amenable to analytical calculations through our mean-field approach, and through simulation we verify the accuracy of our mean-field calculations.

\end{document}